\documentclass{appolb}
\usepackage{graphicx}
\usepackage{amsmath}
\begin{document}
\title{
Studies of the $\omega$ meson with the KLOE detector
%
}
\author{L. Heijkenskj\"old, W. Ikegami Andersson 
\address{on behalf of the KLOE-2 collaboration}
\\
\address{Uppsala University, Department of Physics and Astronomy\\
Box 516, 751 20 Uppsala, Sweden}
}
\maketitle
\begin{abstract}
The paper presents status of three studies involving the $\omega$ meson using data collected by the KLOE detector. The first two projects are feasibility studies performed on simulated data concerning an upper limit measurement of $BR(\phi\to\omega\gamma)$ and the form factor measurement in the $\omega\to\pi^0l^+l^-$ dalitz decay. The third study shows the effect $\pi^0-\pi^0$ interference has in the $\omega\to\pi^+\pi^-\pi^0$ Dalitz plot when $\omega$ is produced through the $e^+e^-\to\omega\pi^0$ channel.

\end{abstract}
\PACS{13.66.Bc, 13.20.Jf, 13.25.Jx}
  
\section{Introduction}
The study of the light mesons is a rich and active field which provides valuable insights into strong interactions taking place at energies where perturbative quantum chromodynamics is no longer valid. Instead effective field theories or dispersion relation calculations can be used. Both of these however need experimental results either as input or as a verification of predictions.

The KLOE detector is situated at the $e^+e^-$ collider DA$\Phi$NE which operates in the $\phi$ meson mass region. The huge available data sets are ideal when searching for the forbidden decay channel $\phi\to\omega\gamma$. The possibility of such a measurement is presented in the first study below.

At these centre of mass energies there is also a large amount of $\omega$ mesons produced, either accompanied by an Initial State Radiation photon or through the $e^+e^-\to\omega\pi^0$ production reaction. In the second study presented below both these production channels are used when estimating how many events of the $\omega\to\pi^0l^+l^-$ decay can be found. In the third study only the $e^+e^-\to\omega\pi^0$ production channel is used to investigate the dynamics of the dominant $\omega$ decay channel, $\omega\to\pi^+\pi^-\pi^0$.

\section{Measuring the C-violating $\boldsymbol{ \phi \rightarrow \omega \gamma}$ branching ratio}
The decay $\phi \rightarrow \omega \gamma$ would violate C-parity in electromagnetic interactions and thus the reaction is not expected to occur. Currently the upper limit of the branching ratio for the reaction is BR$(\phi\rightarrow \omega \gamma) < 5\%$ at 84\% C.L., measured in a bubble chamber experiment performed in the Lawrence Radiation Laboratory in 1966 \cite{Lindsey:1966zz}.

The production reaction $e^+ e^- \rightarrow \omega \gamma_{\mathrm{ISR}}$ has the same final state particles and is the most important background for $\phi \to \omega \gamma$.
With small changes of the centre of mass energy, the change in the cross section for the background reaction $e^+ e^- \rightarrow \omega \gamma_{\mathrm{ISR}}$ should be minimal. However, the cross section of a $\phi \rightarrow \omega \gamma$ decay should vanish when under the $\phi$ meson threshold. Therefore, it becomes plausible to look for the $\phi \rightarrow \omega \gamma$ decay by measuring the $\omega \gamma$ final state cross section around the $\phi$ meson threshold.

In this study, a data sample of $\mathcal{L} = 1.595$ fb$^{-1}$ recorded at $\sqrt{s} = 1019$ MeV/c$^2$ has been used. The $\omega$ meson is detected by looking for its most common decay mode $\omega \rightarrow \pi^+ \pi^- \pi^0$.

The preselection requires at least three neutral energy deposits in the calorimeter not associated to charged tracks; the most energetic one is required to have an energy of $E > 250$ MeV. Additionally two charged tracks with opposite curvature are required to be detected in the drift chamber. After the preselection about $9.7 \cdot 10^6$ events remain in the Monte Carlo sample, where 70302 events originate from $e^+ e^- \rightarrow \omega \gamma_{\mathrm{ISR}}$. 

\begin{figure}[h]
\centering
\includegraphics[scale = 0.27]{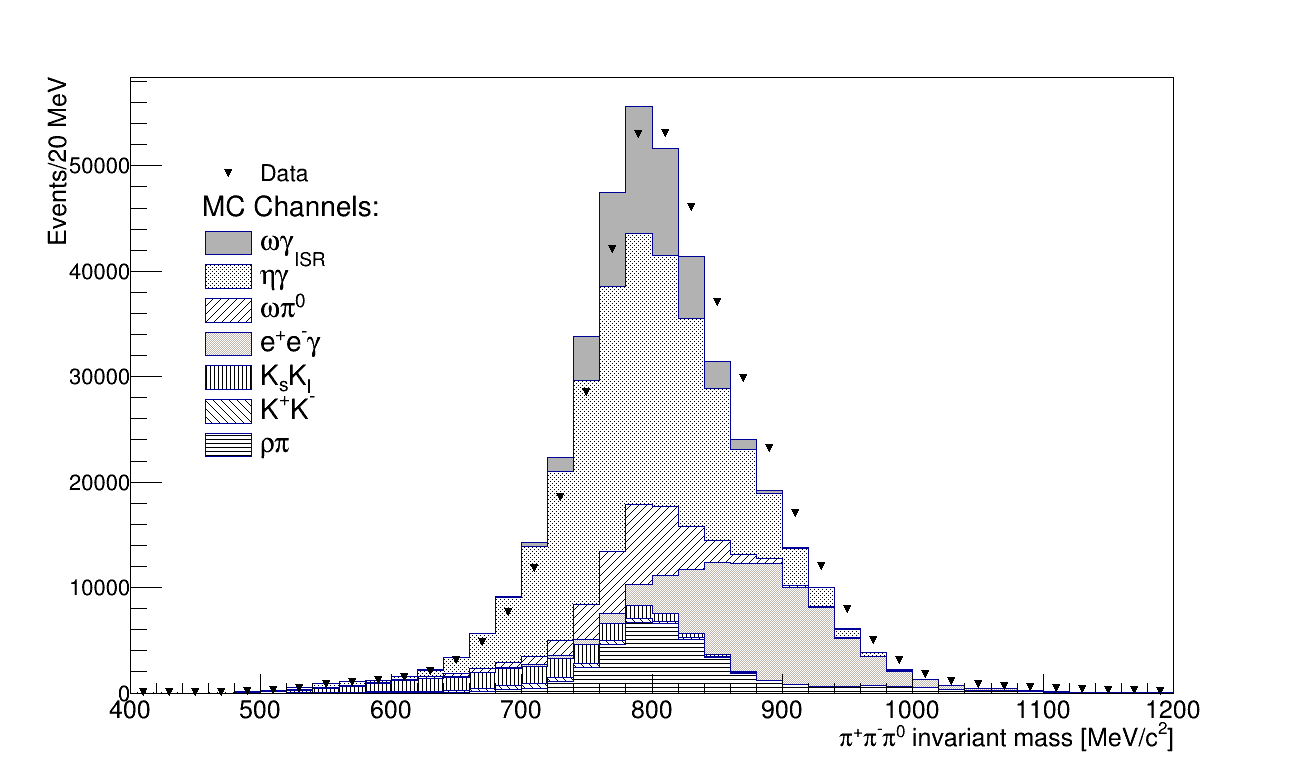}
\caption{Invariant mass of $\pi^+ \pi^- \pi^0$ for data and Monte Carlo channels. All Monte Carlo channels except $\omega \gamma_{\mathrm{ISR}}$, $\omega \pi^0$ and $e^+e^- \gamma$ are $\phi$ resonance channels.}
\label{Mass}
\end{figure}
To suppress non-$\omega \gamma$ background, (i) exactly three photons are required, (ii) the photon pair with invariant mass closest to the $\pi^0$ one is searched for, and (iii) cuts on the energy of the photon recoiling against the omega and on the angle between this photon and the $\omega$ flight direction are applied.


After the background cuts the $\omega \gamma_{\mathrm{ISR}}$ channel has been reduced by 46\%, while the sum of all events have been reduced by 97\%. 

In Fig. \ref{Mass}, the $\pi^+ \pi^- \pi^0$ invariant mass from data is compared to the background expectation from Monte Carlo. Studies are ongoing to further suppress background and find a better agreement between data and Monte Carlo simulations.

\section{$\boldsymbol{\omega \rightarrow \pi^0 l^+ l^-}$ decays in KLOE data}
The magnetic moment of the muon is one of the most precisely measured quantity in particle physics. However, the experimental value deviates from its theoretical prediction \cite{PDG-2012}

Light-meson transition form factors could potentially play a role in the theoretical prediction of the magnetic moment of muons. One of the main models used to predict such form factors is the Vector Meson Dominance model. However, for the $\omega\pi^0$ form factor this model shows a large deviation from the values measured by NA60 experiment in the decay $\omega \rightarrow \pi^0 \mu^+ \mu^-$\cite{Arnaldi}. Alternative theoretical approaches \cite{Terschlusen:2011pm}, attempting to describe the transition form factors of light-meson decays, give better correspondence to the NA60 data and only fails in the higher energy region. 
But on the other hand they are not able to reproduce the recent KLOE preliminary measurement of the form factor of $\phi \to \eta e^+ e^-$. For the solution of this puzzle a second high statistics measurement of the $\omega \pi^0$ form factor would be advisable. Here we present the expected yields for the signal events using existing KLOE data.

A preliminary study has been performed to evaluate the number of $\omega \rightarrow \pi^0 l^+ l^-$ events in the $e^+e^- \rightarrow \omega \gamma_{\mathrm{ISR}}$ and the $e^+e^- \rightarrow \omega \pi^0$ production channels expected in the same data sample of $\mathcal{L} = 1.595$ fb$^{-1}$ used in the previous study.

In addition to taking into account the geometrical acceptance of the KLOE detector, the study also includes the preselection critera: two charged tracks with opposite curvature, at least three charged clusters and one cluster with deposited energy of $E > 250$ MeV.

It is found that $3157 \pm 236$ $\omega \rightarrow \pi^0 e^+ e^-$ decays and $254\pm 74$ $\omega \rightarrow \pi^0 \mu^+ \mu^-$ decays are expected in the KLOE data. These number of events were deemed not satisfactory and more data are needed to perform such a measurement.

\newpage
\section{Dalitz plot studies of $\omega\to\pi^+\pi^-\pi^0$}
The KLOE collaboration has published a measurement of the $e^+e^-\to\omega\pi^0$ production reaction where $1.3\cdot 10^6$ events of $\omega\to\pi^+\pi^-\pi^0$ were found \cite{Kloewpi0}. This sample could in principle be used to study the decay dynamics. A three particle final state decay is studied using a Dalitz plot. The normalised Dalitz plot variables, $X$ and $Y$, are produced as follows, $X=\sqrt{3}\frac{T_{1} - T_{2}}{Q}$ and $Y=\frac{(m_{1} + m_{2} + m_{3})T_{3}}{m_{1}Q}-1$, where $T_i$ is kinetic energy  of the final state particle $i$ in rest frame of the decaying particle and $Q=\sum T_i$. For a distribution which in an $XY$-plot is symmetrically shaped around the centre one can instead use the polar coordinates, $Z=X^2 + Y^2$ and $\Phi = \tan^{-1} \left( \frac{X}{Y}\right)$.

The dynamics of the $\omega\to\pi^+\pi^-\pi^0$ decay includes three main effects. Firstly the three pions are produced predominantly in a P-wave state. This is the prevalent feature of the Dalitz plot distribution which was used to determine the $\omega$ meson quantum numbers \cite{Stevenson}. Secondly the $\rho$ meson plays a role as an intermediate two pion state. Thirdly, any final state $\pi-\pi$ interactions might also affect the density distribution. The last two effects are subtle and have not yet been established since the previous experiments lacked the required statistics. Two recently developed theoretical models have made predictions for this density distribution \cite{Niecknig}\cite{Terschlusen} that could only be tested using a high statistics experimental Dalitz plot distribution, which the data collected by KLOE could provide.

A challenge with the data set presented here consists of the interference caused by interaction between the two final state $\pi^0$. The impact of this interference has been investigated using simulated data and the result will be presented here. In the previous KLOE analysis the process $\omega\pi^0_{1/2}\to\pi^+\pi^-\pi^0_1\pi^0_2$ has been described using the VMD matrix element,
\begin{align}
|\mathbf{J}_{\omega\pi^0}|^2 =& \left| G_{\omega} [ \mathbf{t}_{\omega}(p_{0_1},p_{-},p_{+},p_{0_2}) - \mathbf{t}_{\omega}(p_{0_1},p_{+},p_{-},p_{0_2}) \right. \nonumber\\
 & \left. - \mathbf{t}_{\omega}(p_{0_1},p_{0_2},p_{+},p_{-})] + (p_{0_1} \longleftrightarrow p_{0_2}) \right|^2 \nonumber\\
 =& |\mathbf{J}_{\omega\pi^0_1} + \mathbf{J}_{\omega\pi^0_2}|^2. \label{eq:ME2}
\end{align}
Each $\mathbf{t}_{\omega}$ term accounts for a specific charge of the $\rho$ and permutation of the two final state $\pi^0$'s, for details see \cite{Akhmetshin}. The $\pi^0-\pi^0$ interference is given by the mixed term $\mathbf{J}_{\omega\pi^0_1} \cdot \mathbf{J}_{\omega\pi^0_2}$. To check the magnitude of its influence on the density distribution of the Dalitz plot two data sets were simulated, one with the full matrix element, $|\mathbf{J}_{\omega\pi^0}|^2$, and one where the mixing term was excluded, i.e. only the term $|\mathbf{J}_{\omega\pi^0_1}|^2 + |\mathbf{J}_{\omega\pi^0_2}|^2$ was used. 

\begin{figure}[htb]
\centerline{
\includegraphics[scale=0.6]{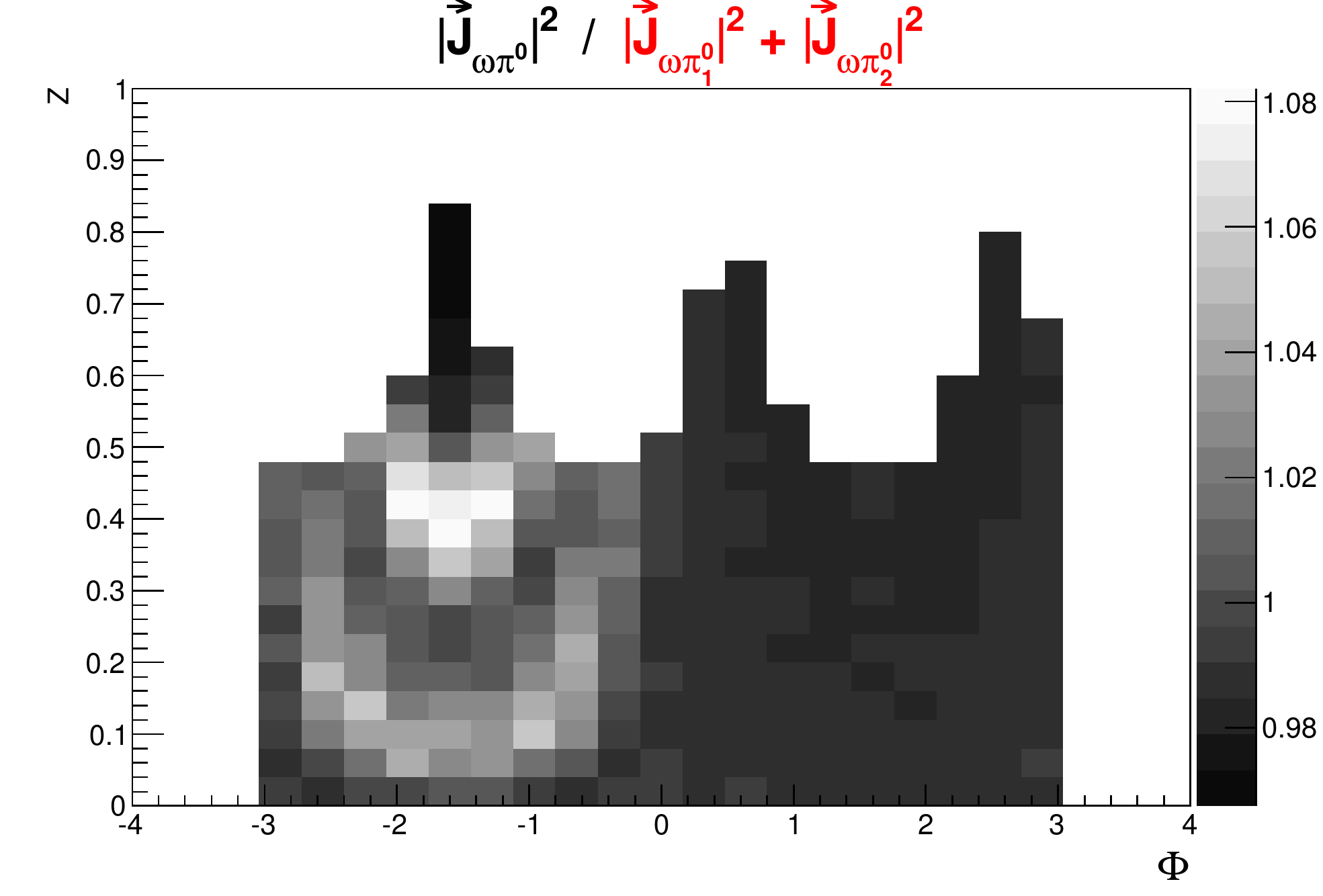}}
\caption{The $Z\Phi$-Dalitz plot density distribution created using the full matrix element, given by equation \eqref{eq:ME2}, is normalised using the density distribution created when the interference term is excluded.}
\label{Fig:w3piDP}
\end{figure}

The main difference between these two matrix elements manifests in the kinematical distributions of the neutral pions. The effect on the density distribution of the $\omega\to\pi^+\pi^-\pi^0$ Dalitz plot is shown in Fig.\ref{Fig:w3piDP}. The  distribution given by the full matrix element is normalised to the distribution given by the matrix element without the mixing term. The maximum deviation reaches $\sim10\%$ for negative values of $\Phi$.
This interference could lead to difficulties in the extraction of the $\omega\to\pi^+\pi^-\pi^0$ Dalitz plot density and subsequently in experimental verification of the theoretical predictions. 

One solution comes from the symmetry of the density distribution in the Dalitz plot which allows for using only part of the plot to predict the full shape. Therefore the minimally disturbed part of the distribution could still be used as an experimental result of the Dalitz plot shape.

It is worth to mention that recently the KLOE-2 detector \cite{kloe2physics} upgraded with new detectors \cite{let,het,kloeIT,CCALT,QCALT} started the data taking campaign and the newly installed Inner Tracker detector \cite{kloeIT} will improve vertex resolution for charged tracks originating from $\omega$ decay.\\

We warmly thank our former KLOE colleagues for the access to the data collected during the KLOE data taking campaign.
We thank the DA$\Phi$NE team for their efforts in maintaining low background running conditions and their collaboration during all data taking. We want to thank our technical staff: 
G.F. Fortugno and F. Sborzacchi for their dedication in ensuring efficient operation of the KLOE computing facilities; 
M. Anelli for his continuous attention to the gas system and detector safety; 
A. Balla, M. Gatta, G. Corradi and G. Papalino for electronics maintenance; 
M. Santoni, G. Paoluzzi and R. Rosellini for general detector support; 
C. Piscitelli for his help during major maintenance periods. 
This work was supported in part by the EU Integrated Infrastructure Initiative Hadron Physics Project under contract number RII3-CT- 2004-506078; by the European Commission under the 7th Framework Programme through the `Research Infrastructures' action of the `Capacities' Programme, Call: FP7-INFRASTRUCTURES-2008-1, Grant Agreement No. 227431; by the Polish National Science Centre through the Grants No. 
DEC-2011/03/N/ST2/02641, 
2011/01/D/ST2/00748,
2011/03/N/ST2/02652,
2013/08/M/ST2/00323,
and by the Foundation for Polish Science through the MPD programme and the project HOMING PLUS BIS/2011-4/3.

\end{document}